\documentclass[3p,times,fleqn]{elsarticle}
\usepackage{amssymb}
\journal{}
\usepackage{amsmath}
\usepackage{amssymb}
\usepackage{psfrag}
\usepackage{color}
\usepackage{stfloats}
\usepackage{fixltx2e}
\usepackage{mathrsfs}
\usepackage{tabularx}
\usepackage{booktabs}
\usepackage{colortbl}
\usepackage{rotating}
\usepackage{boldline}
\usepackage{bm}
\usepackage{changes}
\usepackage{url}
\usepackage{adjustbox}
\usepackage{graphicx}
\usepackage{cellspace}
\usepackage{lineno}

\usepackage{setspace}
\usepackage{moresize}
\usepackage{tabularx}
\usepackage{booktabs}
\usepackage{xcolor}
\usepackage{colortbl}
\usepackage{tikz}
\usetikzlibrary{shapes.geometric, arrows.meta, positioning}
\usepackage{booktabs}
\usepackage{colortbl}
\usepackage{geometry}
\definecolor{lightred}{RGB}{245, 200, 200}   
\definecolor{lightblue}{RGB}{200, 230, 255}  
\definecolor{lightgray}{gray}{0.9}

\usepackage[numbib,notlof,notlot,nottoc]{tocbibind}
\journal{}
\begin{document}
\renewcommand{\topfraction}{0.98}   
\renewcommand{\bottomfraction}{0.98}
\setcounter{topnumber}{3}
\setcounter{bottomnumber}{3}
\setcounter{totalnumber}{4}         
\setcounter{dbltopnumber}{4}        
\renewcommand{\dbltopfraction}{0.98}
\renewcommand{\textfraction}{0.05}  
\renewcommand{\floatpagefraction}{0.95}      
\renewcommand{\dblfloatpagefraction}{0.95}   
\newcommand{\beq}{\begin{equation}}
\newcommand{\eeq}{\end{equation}}
\newcommand{\D}  {\displaystyle}
\newcommand{\DS} {\displaystyle}
\def\sca   #1{\mbox{\rm{#1}}{}}
\def\mat   #1{\mbox{\bf #1}{}}
\def\vec   #1{\mbox{\boldmath $#1$}{}}
\def\scas  #1{\mbox{{\scriptsize{${\rm{#1}}$}}}{}}
\def\scaf  #1{\mbox{{\tiny{${\rm{#1}}$}}}{}}
\def\vecs  #1{\mbox{\boldmath{\scriptsize{$#1$}}}{}}
\def\tens  #1{\mbox{\boldmath{\scriptsize{$#1$}}}{}}
\def\tenf  #1{\mbox{{\sffamily{\bfseries {#1}}}}}
\def\ten   #1{\mbox{\boldmath $#1$}{}}
\sloppy

\tikzstyle{block} = [rectangle, draw=blue!50, thick, minimum height=2em, minimum width=3em, rounded corners=4pt, inner sep=6pt]
\tikzstyle{arrow} = [thick, ->, >=stealth]
\begin{frontmatter}
\title{\Large{\bf{Generalized invariants meet constitutive neural networks:\\ A novel framework for hyperelastic materials}}}

\author[label1]{Denisa Martonová}
\author[label2]{Alain Goriely}
\author[label1,label3]{Ellen Kuhl}
\address[label1]{Institute of Applied Mechanics, Friedrich-Alexander-Universität Erlangen-Nürnberg, Erlangen, Germany}
\address[label2]{Mathematical Institute, University of Oxford, Oxford, United Kingdom}
\address[label3]{Department of Mechanical Engineering, Stanford University, Stanford, CA, United States}
\begin{abstract} 
The major challenge in determining a hyperelastic model for a given material is the choice of invariants and the selection how the strain energy function depends functionally on these invariants. Here we introduce a new data-driven framework that simultaneously discovers appropriate invariants and constitutive models for isotropic incompressible hyperelastic materials. Our approach identifies both the most suitable invariants in a class of generalized invariants and the corresponding strain energy function directly from experimental observations. Unlike previous methods that rely on fixed invariant choices or sequential fitting procedures, our method integrates the discovery process into a single neural network architecture. By looking at a continuous family of possible invariants, the model can flexibly adapt to different material behaviors. We demonstrate the effectiveness of this approach using popular benchmark datasets for rubber and brain tissue. For rubber, the method recovers a stretch-dominated formulation consistent with classical models. For brain tissue, it identifies a formulation sensitive to small stretches, capturing the nonlinear shear response characteristic of soft biological matter. Compared to traditional and neural-network-based models, our framework provides improved predictive accuracy and interpretability across a wide range of deformation states. This unified strategy offers a robust tool for automated and physically meaningful model discovery in hyperelasticity.
\end{abstract}
\end{frontmatter}


\section{Introduction}

\noindent The accurate characterization of hyperelastic materials is essential in both engineering and biomechanics, particularly for materials such as rubber, silicone, and soft biological  tissues. These materials undergo large deformations, and capturing their mechanical response requires constitutive models that accurately describe stress–strain behavior under a wide range of loading conditions. Modeling this behavior is critically important in soft robotics, wearable devices, automotive components, and personalized biomedical simulations, where predictive fidelity under multiaxial stress states is essential.

\noindent The mechanical behavior of hyperelastic solids is typically encoded in a strain energy function \( \psi \) 
whose derivatives can be used to map the deformation gradient onto the local stress state. For isotropic incompressible hyperelastic materials, classical models commonly express the strain energy function \( \psi \) in terms of the first and second invariants \( I_1 \) and \( I_2 \) of the right Cauchy–Green deformation tensor. Several commonly used isotropic hyperelastic models rely on different combinations of these invariants. The neo-Hookean model depends solely on \( I_1 \)~\cite{rivlin1948}, the Blatz-Ko model on \( I_2 \) \cite{blatz_application_1962}, the Mooney–Rivlin model combines \( I_1 \) and \( I_2 \)~\cite{mooney1940}, and the Yeoh model incorporates only \( I_1 \), but uses it in a higher-order polynomial expansion~\cite{yeoh1993}. While these models offer analytical convenience and they are useful for moderate deformations, they often generalize poorly to multiaxial or shear-dominated loading regimes, particularly at large strains~\cite{guo2013,Melly2021}.

\noindent To improve the expressivity and predictive power of constitutive models, recent advances have focused on data-driven approaches, such as neural networks trained to approximate the strain energy function directly from experimental observations. These strategies include constitutive artificial neural networks (CANNs) \cite{linka_constitutive_2021,linka_new_2023,linka_automated_2023-1}, physics-informed neural networks (PINNs) \cite{Raissi2019,Henkes_Wessels_Mahnken_2022}, or physics-augmented neural networks (PANNs) \cite{Linden2023} and encode physical constraints such as material frame indifference throughout the careful selection of \emph{invariant-based} inputs and network architecture. In \emph{principal-stretch-based} models, neural networks use principal stretches as input \cite{pierre_principal-stretch-based_2023,tepole_polyconvex_2025}. This formulation allows for a high degree of flexibility and has proven effective in modeling soft tissues, where mechanical responses are strongly nonlinear. Very recently, material fingerprinting, a novel technology that abandons the use of neural networks and avoids solving nonlinear optimization problems, has been proposed as an effective alternative model discovery \cite{flaschel_material_2025}.

\noindent A notable limitation of these model discovery approaches is the reliance on a fixed invariant or a fixed principal stretch basis. This assumption constrains model expressiveness and may introduce bias, particularly when critical deformation modes are underrepresented in the calibration dataset. 
An  alternative is the use of \emph{generalized invariants} \cite{anssari-benam_generalised_2024}, 
\begin{align}
    && \mathcal{J}_\alpha = \sum_{i=1}^3 \lambda_i^\alpha,
\quad \text{and}\quad \mathcal{K}_\alpha = \prod_{i=1}^3 \lambda_i^\alpha
\end{align}
where the generalized invariants are parameterized in terms of a continuous exponent \( \alpha \in \mathbb R\). This formulation allows for both  compressible materials, for which the set $\{\mathcal{J}_\alpha,\mathcal{J}_{-\alpha},\mathcal{K}_\alpha\}$ forms a complete set of invariants and for incompressible materials, where we can use $\{\mathcal{J}_\alpha,\mathcal{J}_{-\alpha}\}$  and, naturally, $\mathcal{K}_\alpha = I_3 = 1$. In the incompressible case, we recover the classical formulation with \( \alpha = 2 \) yielding the first invariant \( \mathcal{J}_2=I_1 \) and \( \alpha = -2 \) yielding the second invariant \( \mathcal{J}_{-2}=I_2 \). This concept holds the potential to discover new \emph{generalized-invariant–based} models that better match experimental data. A notable widely used example is the Ogden model~\cite{ogden1972large}, which expresses the strain energy function as a weighted sum of powers of the principal stretches, hence a sum of generalized invariants. Indeed, we can rewrite
\begin{align}
    &&\psi = \sum_{j=1}^{n}   \frac{\mu_j}{\alpha_j} \left( \lambda_1^{\alpha_j} + \lambda_2^{\alpha_j} + \lambda_3^{\alpha_j} - 3 \right) = \sum_{j=1}^{n} \frac{\mu_j}{\alpha_j} \left( \mathcal{J}_{\alpha_j} - 3 \right).
\end{align}

\noindent A recent study proposed a two-step procedure to identify the optimal exponent \( \alpha \) and subsequently fit the corresponding strain energy function \citep{anssari-benam_generalised_2024}. While the resulting model is physically interpretable, this strategy demands full and accurate stress–stretch measurements across multiple loading configurations. Moreover, this method requires the strain energy function to be defined \emph{a priori} in terms of the generalized invariants. 
At the same time, another study proposed a comprehensive analysis of how the choice of invariants affects model robustness in PANNs and classical neural network formulations
\cite{dammas_when_2025}. Findings from this study revealed that a model built solely on \( I_1 \) generalizes poorly under multiaxial tests, even if it fits uniaxial data well. These observations align with earlier experimental comparisons~\cite{rosen2019} and underscore the necessity of using both \( I_1 \) and \( I_2 \) to construct stable and predictive constitutive models.

\noindent Despite these insights, important questions remain unresolved: Can we discover constitutive models that depend on {\emph{a single generalized invariant}} and sufficiently explain the data? If such models exist, what is the appropriate form of the generalized invariant, and how does it enter the strain energy definition? If not, what is the {\emph{best set of generalized invariants}} that capture essential features of the material response and allow for a compact formulation of the strain energy function?

\noindent To address these questions, we propose a novel framework that simultaneously discovers the best generalized invariant, the best strain energy format, and the best parameters. Our approach generalizes previous constitutive neural networks and extends \emph{generalized-invariant–based} strategies by making the parameter \( \alpha \) a trainable variable within the neural network. This flexibility allows the network to adaptively explore a continuous space of invariant formulations and fit a physically consistent strain energy function, all directly from experimental stretch–stress data. Importantly, unlike the traditional two-step procedure, our approach discovers the generalized invariant and the format of the strain energy function simultaneously. We validate our method on benchmark data for rubber and brain tissue as examples of engineering and biological soft isotropic materials.

\section{Methods}
\subsection{Generalized invariant framework for constitutive model discovery}
\noindent We frame constitutive model discovery within the theory of finite hyperelasticity. For a deformation described by a smooth mapping \( \varphi: \mathcal{B}_0 \to \mathcal{B} \), the deformation gradient is defined as \( \mathbf{F} = \nabla \varphi \), mapping line elements from the reference configuration \( \mathcal{B}_0 \) to the current configuration \( \mathcal{B} \). The right Cauchy–Green deformation tensor \( \mathbf{C} = \mathbf{F}^\top \cdot \mathbf{F} \) characterizes the local stretch and distortion independent of rigid body motions.
In the following, we consider constitutive models for incompressible isotropic hyperelastic materials. In the classical theory, we express the strain energy function \(\psi\) in terms of the classical invariants of the right Cauchy–Green tensor,  
\begin{equation}
    I_1 = \mathrm{tr}(\mathbf{C})
    \quad \text{and}\quad 
    I_2 = \tfrac{1}{2} \,[ \mathrm{tr}^2(\mathbf{C})  - \mathrm{tr}(\mathbf{C}^2) ].
\end{equation} 
However, recent work introduces a more flexible family of invariant measures 
\cite{anssari-benam_generalised_2024},
\begin{equation}
\mathcal{J}_\alpha = \sum_{i=1}^3 \lambda_i^\alpha,  \label{eq: generalized invariants}
\end{equation}
where \( \lambda_i \) are the principal stretches and \(\alpha\in\mathbb{R}\) is a trainable material parameter. This formulation elegantly generalizes the classical invariants: We recover \(I_1\) for \(\alpha=2\) and \(I_2\) for \(\alpha=-2\). 

\subsubsection{Two-step approach}

\noindent The initial method relies on a two-step approach that {\emph{first}} identifies an optimal exponent $\alpha$ in (\ref{eq: generalized invariants}) and {\emph{then}} fits a specific strain energy function $\psi=\psi(\mathcal{J}_\alpha)$ to the data \citep{anssari-benam_generalised_2024}. In the first step, an optimal generalized invariant exponent $\alpha$ is found independently of any specific choice of strain energy function $\psi$ via so-called pseudo‑universal relationships. This is accomplished by leveraging the fact that, under certain loading conditions, such as biaxial extension, a strain-energy-independent stress quotient can be formulated when the strain energy function depends on a single invariant. To illustrate this strategy, we start from the general expression for the Cauchy stress tensor $\boldsymbol{\sigma}$ in terms of a generalized invariant $\mathcal{J}_\alpha$,
\begin{equation}
    \boldsymbol{\sigma} = \frac{\partial \psi}{\partial \mathcal{J}_\alpha} \frac{\partial \mathcal{J}_\alpha}{\partial \mathbf{F}} \cdot \mathbf{F}^{\top} - p \,\mathbf{I} \,, \label{eq: cauchy stress}
\end{equation}

\noindent where \( p \) is the hydrostatic pressure
and $\mathbf{I}$ is the identity tensor. In the specific case of a biaxial extension with stretches $\lambda_1$ and $\lambda_2$ in two orthogonal directions and zero traction in the out-of-plane direction, $\sigma_{33} = 0$, the deformation gradient takes the following explicit form
\begin{equation}
    \mathbf{F} = \text{diag} \{\lambda_1, \lambda_2, (\lambda_1 \lambda_2)^{-1}\},
\end{equation}
based on the plane-stress assumption. Under these loading conditions, the principal components of the Cauchy stress tensor become
\begin{align}
    \sigma_{11} &=  \frac{\partial \psi}{\partial \mathcal{J}_\alpha} \alpha \,[  \lambda_1^\alpha - (\lambda_1 \lambda_2)^{-\alpha} \,]
    \quad \mbox{and} \quad
    \sigma_{22} =  \frac{\partial \psi}{\partial \mathcal{J}_\alpha} \alpha \, [ \lambda_2^\alpha - (\lambda_1 \lambda_2)^{-\alpha} \,]. \label{eq: sigma22}
\end{align}
In the ratio of the two stress components, the unknown derivative $\partial \psi / \partial \mathcal{J}_\alpha$ cancels, and the quotient depends only on $\alpha$ and the stretches $\lambda_1$ and $\lambda_2$,
\begin{equation}
    \frac{\sigma_{11}}{\sigma_{22}} = \frac{  \lambda_1^\alpha - (\lambda_1 \lambda_2)^{-\alpha} }{  \lambda_2^\alpha - (\lambda_1 \lambda_2)^{-\alpha} }. \label{eq: stress quotient}
\end{equation} 
\noindent Alternatively, we could use the chain rule by first computing the derivatives of the generalized invariants with respect to the classical first and second invariants \cite{anssari-benam_generalised_2024}. The expression in (\ref{eq: stress quotient}) is entirely independent of the form of the strain energy function, making it suitable to directly estimate the power $\alpha$. For a dataset of measured stress components 
${\, \hat{\sigma}_{11}, \hat{\sigma}_{22}\,}$ and corresponding stretches 
${\, \lambda_1, \lambda_2 \,}$, the loss function is defined as
\begin{equation}
    L_1(\alpha; \mathbf{F}) = \frac{1}{n_{\text{data}}} \sum_{i=1}^{n_{\text{data}}} \left\| \frac{\sigma_{11}(\mathbf{F}_i, \alpha)}{\sigma_{22}(\mathbf{F}_i, \alpha)} - \frac{\hat{\sigma}_{11}}{\hat{\sigma}_{22}} \right\|^2_2 \quad \longrightarrow \quad \min.
\end{equation}
Minimizing this loss with respect to $\alpha$ yields the optimal exponent $\alpha^*$ that captures the material's generalized invariance behavior based purely on experimental stress data. Importantly, this identification of $\alpha$ is performed without {\emph{any}} assumption about the specific form of the strain energy function $\psi$. 
In the second step, the optimal exponent \(\alpha^*\) is identified and the stress quotient method in the first step is used to select and fit arbitrary stress-strain data to a closed‑form strain energy function \( \psi \) that \textit{only depends}  on the generalized invariant $\mathcal{J}_{\alpha^*} = \sum_{i=1}^3 \lambda_i^{\alpha^*}$. There are two possible choices \cite{anssari-benam_generalised_2024},
the {\emph{one-term Ogden‑type strain energy}} \cite{ogden1972large},
\begin{equation}
    \psi^{\mathrm{OG}} =  \frac{\mu_j}{\alpha^*} \left( \mathcal{J}_{\alpha^*} - 3 \right), \label{eq: psi OG}
\end{equation}
where \( \mu > 0 \) is a stress-like material constant, and
the {\emph{one-term Anssari‑Benam strain energy}} \cite{anssari-benam_large_2023},
\begin{equation}
    \psi^{\mathrm{AB}} = \frac{3(n - 1)}{2n} \mu N \left[ 
    \frac{1}{3N(n - 1)} \left(\mathcal{J}_{\alpha^*} - 3 \right)
    - \ln \left( \frac{\mathcal{J}_{\alpha^*} - 3N}{3 - 3N} \right)
    \right], \label{eq: psi AB}
\end{equation}
where $n > 0$ and $N > 0$ are additional dimensionless model parameters. Notably, when  $N \to \infty$, we recover the one-term Ogden model in (\ref{eq: psi OG}).
Given a fixed $\alpha^*$ from the first step, we identify 
the model parameters 
$\mathbf{w} = \{ \, \mu \, \}$ or 
$\mathbf{w} = \{ \, n, N \, \}$, by minimizing the loss function, 
\begin{equation}
    L_2(\mathbf{w}; \mathcal{J}_{\alpha^*}(\mathbf{F})) = \frac{1}{n_{\text{data}}} \sum_{i=1}^{n_{\text{data}}} 
    \left\| \, \boldsymbol{\sigma}^{\alpha}(\mathcal{J}_{\alpha^*}(\mathbf{F}_i), \mathbf{w}) - \hat{\boldsymbol{\sigma}}_i \, \right\|_2^2 
    \; \rightarrow \; \min. \label{eq loss step 2}
\end{equation}
where $\boldsymbol{\sigma}^{\alpha}$ are the modeled Cauchy stresses according to (\ref{eq: cauchy stress}) for a specific strain energy function $\psi^{\mathrm{OG}}$ in (\ref{eq: psi OG})  or $\psi^{\mathrm{AB}}$ in (\ref{eq: psi AB}) and $\hat{\boldsymbol{\sigma}}_i$ denote the measured data.

\subsection{Constitutive neural networks for  model discovery}
\begin{figure}[h!]
\centering
\includegraphics[width=\textwidth]{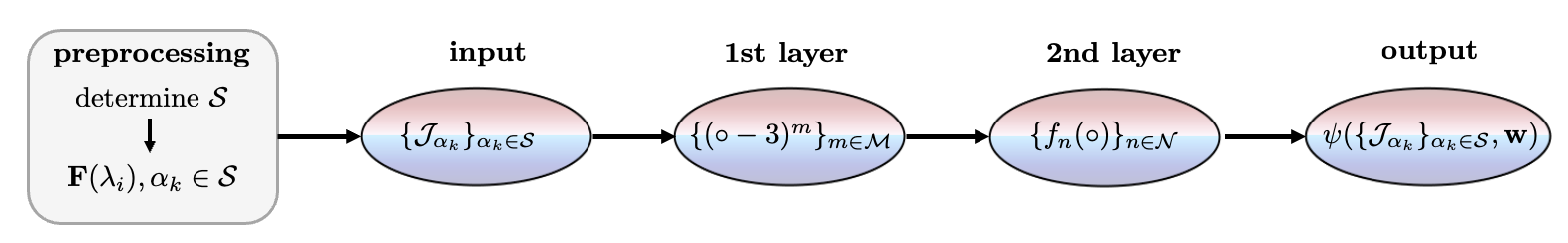}
\caption{GI-CANN network architecture, where $\mathcal{S}$ is the set of all exponents $\alpha_k$ of stretches $\lambda_i$;
$\mathcal{M}$ is the set of all powers of generalized invariants $\mathcal J_{\alpha_k}$;
$\mathcal{N}$ is the set of all functions applied to powers of generalized invariants;
$\mathbf{w}$ are all network weights or model parameters.}
\label{fig: GI-CANN}
\end{figure}

\noindent To circumvent the need for a predefined strain energy function form and a sequential two-step identification procedure, we propose the \emph{generalized-invariant-based} CANN (GI-CANN), which enables simultaneous learning of both, the optimal input invariants and the corresponding strain energy function directly from stress-strain data measured during arbitrary deformations. Figure~\ref{fig: GI-CANN} sketches the generalized architecture of the GI‑CANN proposed in this study. 

\noindent In a preprocessing step, we specify a set of exponents \( \mathcal{S}\subset\mathbb{R} \) that are either fixed or that the network discovers during training. From the measured deformation modes, characterized by the deformation gradient $\mathbf F$, we then evaluate the generalized invariants \( \mathcal{J}_{\alpha_k}=\sum_{i=1}^{3}\lambda_i^{\alpha_k} \) for every \( \alpha_k\in\mathcal{S} \).  The first network layer ensures a stress-free reference state by first subtracting the trace of the identity tensor, $\mathrm{tr}(\mathbf{I})=3$, and then applying powers to these corrected invariants. The second layer subjects these quantities to convex functions \( \rm f_n \) with \(n\!\in\!\mathcal{N} \), for example the identity $\rm f_{\mathrm{id}}(\circ)=\circ$, the natural logarithm $\rm f_{\ln} = \ln(1-\circ)$, or the exponential function $\rm f_{\exp}=\exp(\circ)-1$. Subsequently, the layers combine the transformed features and yield the generalized strain energy function,
\begin{equation} \label{eq:general free energy}
\psi = \sum_{\alpha_k \in \mathcal{S}} 
   {\rm w}^{\rm{id}}_{k} \left[ \mathcal{J}_{\alpha_k} - 3 \right]
 + {\rm w}^{\rm{exp}}_{k,2} \left[ \exp \left( {\rm w}^{\rm{exp}}_{k,1} \left[ \mathcal{J}_{\alpha_k} - 3 \right]\right) - 1 \right]
 + {\rm w}^{\rm{ln}}_{k,2} \left[ \ln\left(1 - {\rm w}^{\rm{ln}}_{k,1} \left[\mathcal{J}_{\alpha_k} - 3\right]\right) \right] \,.
\end{equation} 
\noindent Here, we treat both the weights $ \mathbf{w} = \{\rm w^{\rm{id}}_{k}, \rm w^{\rm{exp}}_{k,1}, \rm w^{\rm{exp}}_{k,2}, \rm w^{\rm{ln}}_{k,1}, \rm w^{\rm{ln}}_{k,2} \}$ and the exponents \( \mathcal S = \{\alpha_k\} \) as trainable parameters, subject to the following loss function,
\begin{equation}
L(\alpha,\mathbf{w}; \mathbf{F}) = \frac{1}{n_{\text{data}}} \sum_{i=1}^{n_{\text{data}}} 
    \left\| \, \boldsymbol{\sigma}^{\alpha}(\mathcal{J}_{\alpha}(\mathbf{F}_i), \mathbf{w}) - \hat{\boldsymbol{\sigma}}_i \, \right\|_2^2  + r \, \sum_{i=1}^{n_{\rm{weights}}} | \, w_i \, |
    \; \rightarrow \; \min, \label{eq: loss GI-CANN}
\end{equation}
\noindent We add an \(L_1\)-type regularization term, \(r \, \| \, \mathbf{w}\,  \|_1\), to our loss function to induce model sparsity and increase interpretability~\citep{mcculloch_sparse_2024}. In the following, we set $r=0.01$~\cite{martonova_discovering_2025,martonova_automated_2024, vervenne_constitutive_2025}.
By learning the exponent \(\alpha\) directly from the data, the GI‑CANN eliminates the sequential procedure of the two‑step approach~\citep{anssari-benam_generalised_2024} and dispenses with any \emph{a priori} assumption on the form of the strain energy function.
Moreover, the GI‑CANN subsumes earlier constitutive neural network approches as special cases. If we restrict the exponent set to the classical values, 
\(\mathcal{S}=\{-2,2\}\), the architecture reduces to the \emph{standard-invariant-based} CANN (SI‑CANN)~\citep{linka_new_2023,linka_automated_2023-1}. Conversely, prescribing a large discrete grid of integer exponents, \(\mathcal{S}=\{-n,\dots,n\}\), while exclusively activating $\rm{f_{id}}$ in the second layer recovers the \emph{principal-stretch-based} CANN (PS‑CANN)~\citep{pierre_principal-stretch-based_2023}. Table~\ref{tab:model_comparison} summarizes the key differences between these architectures and highlights how our new GI‑CANN integrates and generalizes both by unifying invariant selection and strain energy function identification in a single, data‑driven optimization procedure. The GI-CANN naturally provides a flexible framework that generalizes well to common isotropic constitutive models by expressing their strain energy functions as specific cases of the strain energy function in (\ref{eq:general free energy}), see Table~\ref{tab:models_prototype} for an overview.

\begin{table}[ht]
\centering
\caption{Overview of isotropic and incompressible constitutive neural network  types. The table contrasts how each model type determines its invariant feature set \( \mathcal{S} \), processes these invariants as input, applies transformations in layered neural architectures, and produces the final form of the strain energy function \( \psi \). Our proposed method (bottom row) learns the generalized exponent \( \alpha \) and strain energy function \( \psi \) simultaneously, enabling autonomous model discovery.}
\scriptsize\renewcommand{\arraystretch}{1.4}
\begin{tabularx}{\textwidth}{>{\raggedright\arraybackslash}p{3.cm} >{\raggedright\arraybackslash}p{2.2cm} >{\raggedright\arraybackslash}p{2.2cm} >{\raggedright\arraybackslash}p{2.cm} >{\raggedright\arraybackslash}p{2.cm} >{\raggedright\arraybackslash}X}
\toprule
\rowcolor{lightgray}
\textbf{isotropic incompressible CANN / model type} & \textbf{\( \mathcal{S} \) determination} & \textbf{input (generalized invariants)} & \textbf{1st layer \( \mathcal{M} \) (powers)} & \textbf{2nd layer \( \mathcal{N} \) (functions)} & \textbf{output (strain energy function)} \\
\midrule

SI-CANN \cite{linka_new_2023}  & 
\emph{a priori} given \newline \( \mathcal{S} = \{-2, 2\} \) & 
\( \mathcal{J}_2 = I_1 \), \newline \( \mathcal{J}_{-2} = I_2 \) & 
\( \{1, 2\} \) & 
\(  \{\mathrm{id}, \exp, \ln\} \) & 
\( \psi(\mathcal{J}_2, \mathcal{J}_{-2}, \mathbf{w}) \) \\

PS-CANN \cite{pierre_principal-stretch-based_2023} & 
\emph{a priori} given \newline \( \mathcal{S} = \{-n, \ldots, n\} \) & 
\( \mathcal{J}_{-30}, \ldots, \mathcal{J}_{10} \) & 
\(  \{1\} \) & 
\(  \{\mathrm{id}\} \) & 
\( \psi(\mathcal{J}_{-n}, \ldots, \mathcal{J}_{n}, \mathbf{w}) \) \\

two-step approach \cite{anssari-benam_generalised_2024} & 
via minimization in step 1, \( \mathcal{S} = \{\alpha^*\} \) & 
\( \mathcal{J}_{\alpha^*} \) & 
\(  \{1\} \) & 
\(  \{\mathrm{id}, f_\mathrm{AB}\} \) & 
\( \psi^{\mathrm{OG}}(\mathcal{J}_{\alpha^*}) \) or \( \psi^{\mathrm{AB}}(\mathcal{J}_{\alpha^*}) \) \\

GI-CANN & 
\( \alpha \) trainable parameter \newline \( \mathcal{S} = \mathbb{R} \) & 
\( \{\mathcal{J}_{\alpha_k}\}_{\alpha_k \in \mathbb{R}} \) & 
\(  \{1\} \) & 
\(  \{\mathrm{id}, \exp, \ln\} \) & 
\( \psi(\mathcal{J}_{\alpha^*}, \mathbf{w}, \alpha^*) \) \\

\bottomrule
\end{tabularx}
\label{tab:model_comparison}
\end{table}

\begin{table}[ht]
\centering
\caption{Common isotropic incompressible constitutive models expressed as specific cases of GI-CANN strain energy function, see Eq.~(\ref{eq:general free energy}).}
\scriptsize\renewcommand{\arraystretch}{1.4}
\begin{tabularx}{\textwidth}{>{\raggedright\arraybackslash}X >{\raggedright\arraybackslash}X >{\centering\arraybackslash}X >{\centering\arraybackslash}X >{\centering\arraybackslash}X}
\toprule
\rowcolor{lightgray}
\textbf{model} & \textbf{strain energy function \(\psi\)} & \textbf{exponents} \(\mathcal{S}\)  & \textbf{powers} \(\mathcal{M}\)  & \textbf{functions} \(\mathcal{N}\)  \\
\midrule

Blatz-Ko & 
\(\mathrm{w}_1^{\mathrm{id}} [\mathcal{J}_{-2} - 3]\) & 
\(\{-2\}\) & 
\(\{1\}\) & 
id \\

Demiray & 
\(\mathrm{w}_1^{\mathrm{exp}} \left[\exp\left(\mathrm{w}_2^{\mathrm{exp}} [\mathcal{J}_2 - 3]\right) - 1\right]\) & 
\(\{2\}\) & 
\(\{1\}\) & 
exp \\

Gent & 
\(\mathrm{w}_1^{\mathrm{ln}} \ln\left(1 - \mathrm{w}_2^{\mathrm{ln}} [\mathcal{J}_2 - 3]\right)\) & 
\(\{2\}\) & 
\(\{1\}\) & 
ln \\

Holzapfel & 
\(\mathrm{w}_1^{\mathrm{exp}} \left[\exp\left(\mathrm{w}_2^{\mathrm{exp}} [\mathcal{J}_2 - 3]^2\right) - 1\right]\) & 
\(\{2\}\) & 
\(\{2\}\) & 
exp \\

Mooney-Rivlin & 
\(\mathrm{w}_1^{\mathrm{id}} [\mathcal{J}_2 - 3] + \mathrm{w}_2^{\mathrm{id}} [\mathcal{J}_{-2} - 3]\) & 
\(\{2, -2\}\) & 
\(\{1\}\) & 
id \\

neo-Hookean & 
\(\mathrm{w}_1^{\mathrm{id}} [\mathcal{J}_2 - 3]\) & 
\(\{2\}\) & 
\(\{1\}\) & 
id \\

one-term Ogden & 
\(\mathrm{w}_1^{\mathrm{id}} [\mathcal{J}_\alpha - 3]\) & 
\(\{\alpha\}\) & 
\(\{1\}\) & 
id \\

\bottomrule
\end{tabularx}
\label{tab:models_prototype}
\end{table}

\subsection{Benchmarking}
\noindent We demonstrate the features of the proposed approach on two representative experimentally measured datasets for isotropic incompressible materials, rubber  \cite{treloar_stress-strain_1944} and brain \cite{budday_mechanical_2017}, see Table \ref{tab:rubber_brain_data} for the specific stress-strain data used in our work. For the rubber dataset, we train the network using uniaxial and equibiaxial extension data and test it on pure shear data. For the brain dataset, we train the model on uniaxial tension and compression data and test it on simple shear data. 
The generalized invariants and Cauchy stress have the following formats. 
For uniaxial extension or compression with the principal stretches \(\lambda_1\) and \(\lambda_2 = \lambda_3 = \lambda_1^{1/2}\), we obtain
\begin{alignat}{3}
\mathcal{J}_\alpha &= \lambda_1^{\alpha} + 2 \lambda_1^{-\alpha/2} 
&\quad &\rm{and} &\quad
&\sigma_{11} = \alpha \frac{\partial \psi}{\partial \mathcal{J}_\alpha} \left( \lambda_1^\alpha - \lambda_1^{-\alpha/2} \right).
\label{eq:sigma11_uni}
\end{alignat}
\noindent In the equibiaxial extension where \(\lambda_1 = \lambda_2\) and $\lambda_3=(\lambda_1\lambda_2)^{-1}$, Eqs. (\ref{eq:sigma11_uni}) and (\ref{eq: sigma22}) simplify to
\begin{alignat}{3}
\mathcal{J}_\alpha &= 2 \lambda_1^{\alpha} + \lambda_1^{-2\alpha} 
&\quad &\rm{and} &\quad
&\sigma_{11} = \sigma_{22} = \alpha \frac{\partial \psi}{\partial \mathcal{J}_\alpha} \left( \lambda_1^\alpha - \lambda_1^{-2\alpha} \right),
\label{eq:sigma11_equibiax}
\end{alignat}
\noindent whereas for pure shear deformation where \(\lambda_2 = 1\) and $\lambda_3=\lambda_1^{-1}$, we obtain
\begin{alignat}{3}
\mathcal{J}_\alpha &= \lambda_1^{\alpha} + 1 + \lambda_1^{-\alpha} 
&\quad &\rm{and} &\quad
&\sigma_{11} = \alpha \frac{\partial\psi}{\partial\mathcal{J}_\alpha} \left( \lambda_1^\alpha - \lambda_1^{-\alpha} \right).
\label{eq:sigma11_ps}
\end{alignat}
\noindent Finally, the simple shear test with shear strain \(\gamma\) results in
\begin{alignat}{3}
\mathcal{J}_\alpha &= 2 \sqrt{1 + \frac{\gamma^2}{4}} + 1 
&\quad &\rm{and} &\quad
&\sigma_{12} = \frac{\alpha}{2^{\alpha}} \frac{\partial \psi}{\partial \mathcal{J}_\alpha}\frac{\left( \gamma + \sqrt{\gamma^2 + 4} \right)^{\alpha} - \left( -\gamma + \sqrt{\gamma^2 + 4} \right)^{\alpha}}{\sqrt{\gamma^2 + 4}}.
\label{eq:sigma12_ss}
\end{alignat}
\noindent For both materials, rubber and brain, we conduct model discovery in a staged manner to identify the most relevant generalized invariants. We begin by restricting the strain energy function to depend on a single generalized invariant \( \mathcal{J}_{\alpha_+} \) with a positive exponent \( \alpha_+ \), corresponding to an \( I_1 \)-like formulation, represented by reddish colors in the visualizations. In a second discovery run, we allow the model to use only a single invariant with a negative exponent \( \alpha_-  \), yielding an \( I_2 \)-like formulation, represented by bluish colors in the visualizations. Finally, we perform model discovery with two generalized invariants simultaneously, one with \( \alpha_+ \) and one with \( \alpha_- \), allowing the network to learn a mixed formulation that captures both types of mechanical response.~This progression enables us to systematically uncover the generalized invariant  most suitable for each material. \\[4.pt]
\noindent To evaluate model performance, we compare the computationally predicted and experimentally measured responses using the coefficient of determination \(R^2\), which quantifies the goodness of fit. We assess and compare the goodness of fit across various approaches, including SI-CANN, PS-CANN and our proposed GI-CANN as well as results from \cite{dammas_when_2025} comparing unconstrained feedforward neural networks (FNN) and fully input convex neural networks (FICNN) using the same training and testing data. We minimize the loss function~(\ref{eq: loss GI-CANN}) using the Adam algorithm~\cite{kingma_adam_2017}.

\section{Results and Discussion}

\begin{figure}[h]
    \centering
\includegraphics[width=\textwidth]{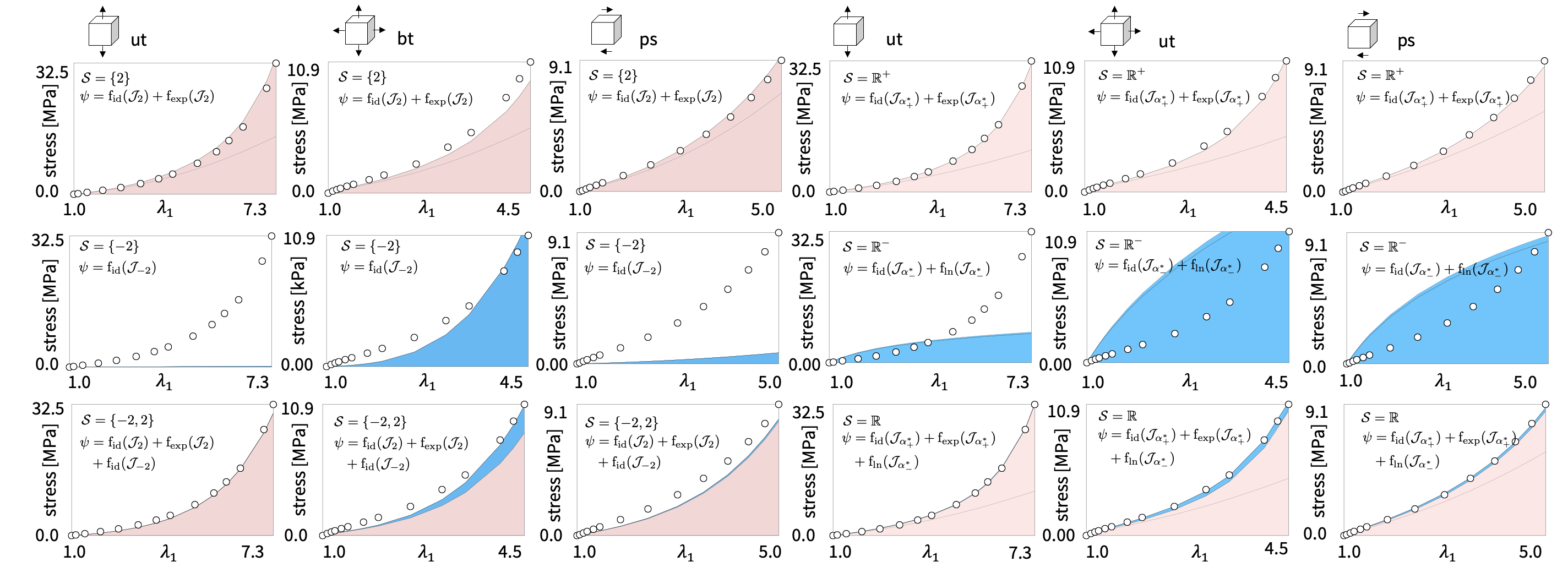}
     \caption{Stress-stretch response of rubber. Training under uniaxial tension (ut) and equibiaxial tension (bt), and testing under pure shear (ps) yields different  strain energy functions $\psi$. Each column represents a loading mode (ut, bt, ps), and each row corresponds to a different set of input generalized invariants $\mathcal{S}$. Experimental data are shown as white circles, and colored regions indicate the contributions of the specific terms to the overall modeled Cauchy stress components. The color scale reflects the influence of the parameter $\alpha$ used in the definition of $J_\alpha$, with blue for negative $\alpha$, red for positive $\alpha$, and lighter tones indicating smaller values, see legend in Figure \ref{fig:brain-plots}.}
    \label{fig:rubber-plots}
\end{figure}

\begin{figure}[h]
    \centering
    \includegraphics[width=\textwidth]{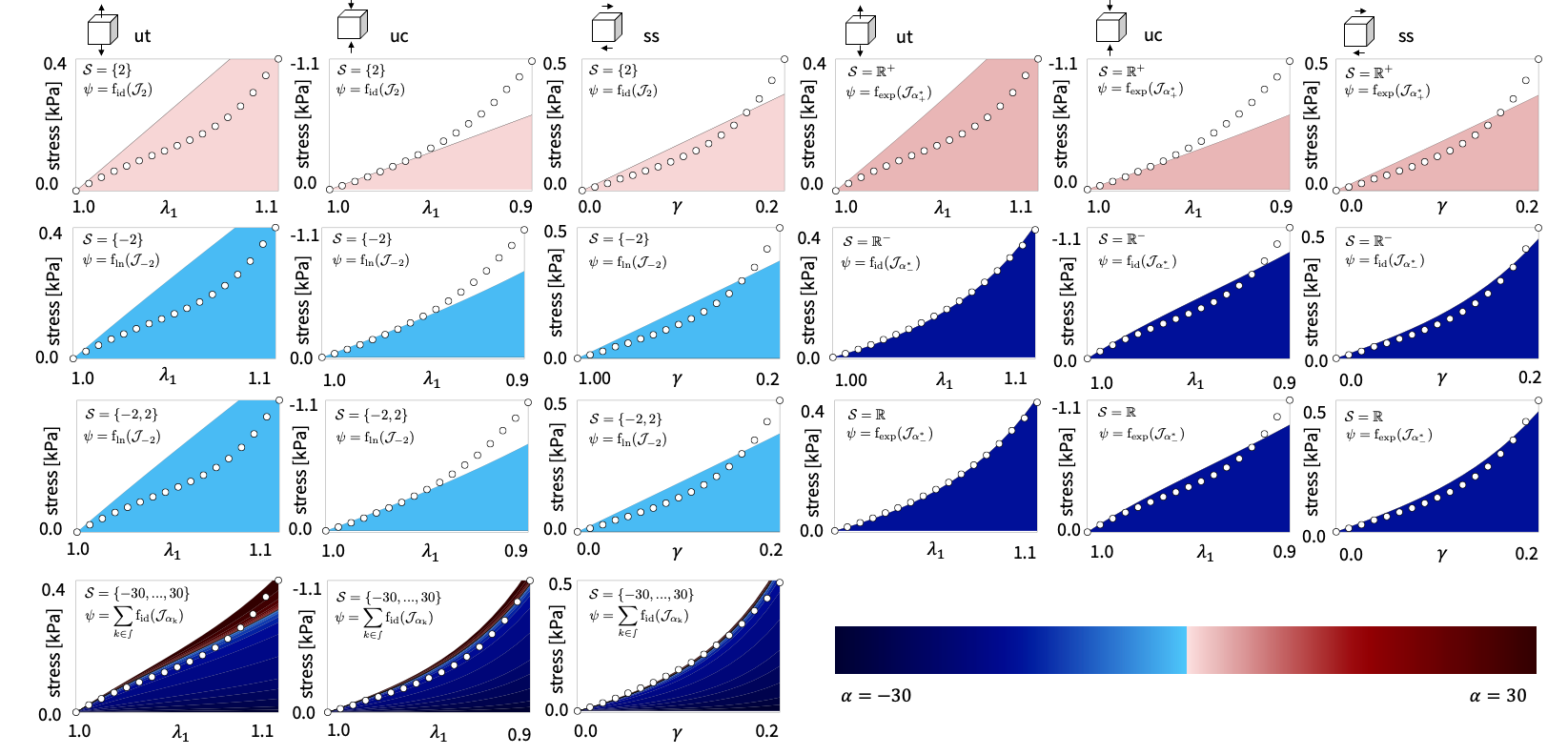}
    \caption{Stress-stretch response of brain tissue. Trained under uniaxial tension (ut) and uniaxial compression (uc), and testing under simple shear (ss) yields different  strain energy functions $\psi$. Each column represents a loading mode (ut, uc, ss), and each row corresponds to a different set of input generalized invariants $\mathcal{S}$. Experimental data are shown as white circles, and colored regions indicate the contributions of the specific terms to the overall modeled Cauchy stress components. The color scale reflects the influence of the parameter $\alpha$ used in the definition of $J_\alpha$, with blue for negative $\alpha$, red for positive $\alpha$, and lighter tones indicating smaller values.}
    \label{fig:brain-plots}
\end{figure}

\begin{figure}[t]
    \centering
\includegraphics[width=\textwidth]{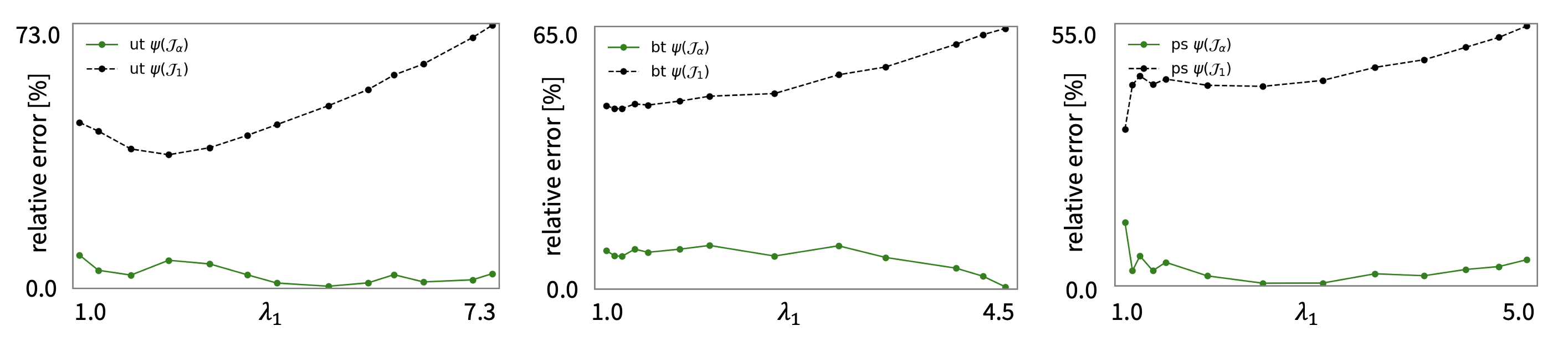}
 \caption{Relative error plots for rubber models. Comparison between the generalized-invariant-based model $\psi(\mathcal{J}_\alpha)$ in (\ref{eq: best model rubber}) and a model $\psi(I_1)$ based on the standard invariant $I_1$. The relative error is computed as $\left|\frac{\sigma_{ij}^\alpha - \hat{\sigma}_{ij}}{\hat{\sigma}_{ij}}\right| \times 100$, where $\sigma_{ij}^\alpha$ and $\hat{\sigma}_{ij}$ denote the modeled and measured Cauchy stress components, respectively. From left to right: uniaxial tension (ut), equibiaxial tension (bt), and pure shear (ps).}
\label{fig:error-plots-rubber}

\end{figure}

\begin{figure}[t]
    \centering
\includegraphics[width=\textwidth]{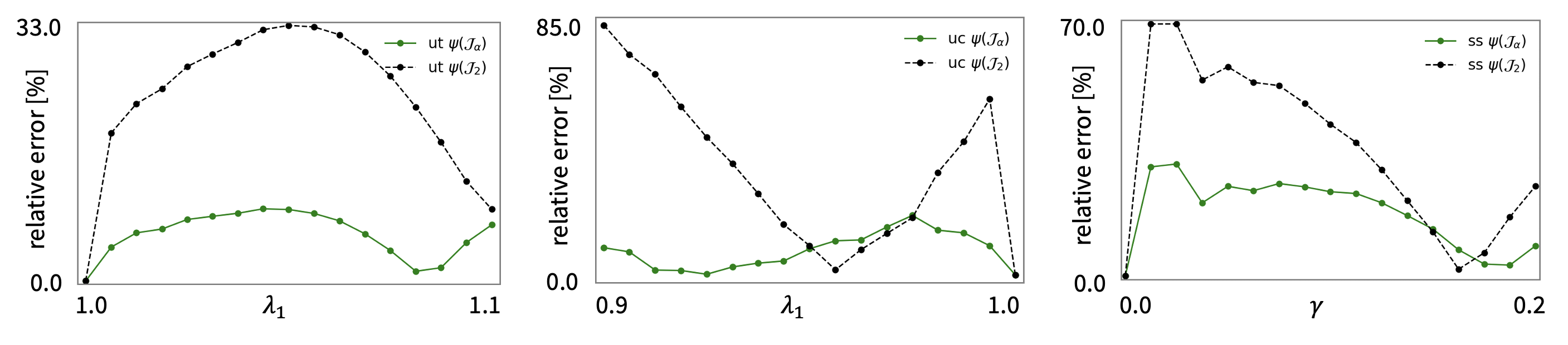}
  \caption{Relative error plots for brain cortex models. Comparison between the generalized-invariant-based model $\psi(\mathcal{J}_\alpha)$ in (\ref{eq: best brain model}) and a model $\psi(I_2)$ based on the standard invariant $I_2$. The relative error is computed as $\left|\frac{\sigma_{ij}^\alpha - \hat{\sigma}_{ij}}{\hat{\sigma}_{ij}}\right| \times 100$, where $\sigma_{ij}^\alpha$ and $\hat{\sigma}_{ij}$ denote the modeled and measured Cauchy stress components, respectively. From left to right: uniaxial tension (ut), uniaxial compression (uc), and simple shear (ss).}
\label{fig:error-plots-brain}

\end{figure}

\noindent Across both rubber and human brain tissue, models based on generalized invariants $\mathcal{J}_\alpha$ consistently outperform traditional approaches that rely solely on classical invariants such as $I_1$ or $I_2$. This performance advantage is apparent in the stress response plots in Figures~\ref{fig:rubber-plots} and~\ref{fig:brain-plots}, and is quantitatively confirmed by the accuracy metrics  in Tables~\ref{tab:rubber-results} and~\ref{tab:brain-results}. The specific model parameters for rubber and brain are reported in Tables \ref{tab:rubber_parameters} and \ref{tab:brain_parameters}.\\[4.pt]
\noindent For the rubber data, the mechanical response is best described by models with {\emph{positive exponents}} $\alpha_+$. These values correspond to an $I_1$-type behavior, and models based on this formulation yield significantly higher fidelity in reproducing experimental stress-stretch data. In particular, the best model 
\begin{equation}
    \psi = \mathrm{f_{id}}(\mathcal{J}_{1.675}) + \mathrm{f_{exp}}(\mathcal{J}_{1.675}) = 0.227\,\mathrm{MPa} \cdot (\mathcal{J}_{1.675} - 3) + 0.297\,\mathrm{MPa} \cdot \left[\exp(0.107 \cdot (\mathcal{J}_{1.675} - 3)) - 1\right]. \label{eq: best model rubber}
\end{equation}
combines an identity function and an exponential term, both formulated on the same generalized invariant $\mathcal{J}_{\alpha_+^*}$ with $\alpha_+^* = 1.675$. 
This formulation results in an excellent fit across all deformation modes, achieving an average coefficient of determination $R^2 = 0.998$, as detailed in Table~\ref{tab:rubber-results}. The visual evidence of this accuracy is confirmed by the first row of Figure~\ref{fig:rubber-plots}, particularly in columns three through six, where the modeled responses closely align with the experimental data shown as white circles. Red-shaded regions dominate the visualizations, indicating a strong influence of positive $\alpha_+$ terms on the overall Cauchy stress components. These regions highlight the model's reliance on $I_1$-like effects.
\noindent The relative error plots  in Figure~\ref{fig:error-plots-rubber} further confirm the accuracy of the discovered model in (\ref{eq: best model rubber}), which  outperforms the discovered \emph{standard-invariant-based} model, $\psi(I_1)=\rm{f_{id}}(I_1) + \rm{f_{exp}}(I_1)$, with the best performance achieved under uniaxial tension. Compared to the error plots of the Anssari-Benam model (\ref{eq: psi AB}), our discovered model produces smaller errors at low stretches but larger errors at high stretches. Nevertheless, the strength of the proposed approach lies in its ability to automatically identify a suitable model rather than relying on \emph{a priori} assumptions about the strain energy function.
 We can further improve the model predictions by including {\emph{both positive and negative exponents}} $\alpha$. In these cases, the regularization embedded in the GI-CANN architecture selectively activates only those components that contribute meaningfully to the stress response. This selective activation is particularly evident in models that use the combined invariant set $\{\mathcal{J}_{\alpha_+}, \mathcal{J}_{\alpha_-}\}$. The two-invariant model achieves a near-perfect fit $R^2 = 0.999$, as shown in the bottom section of Table~\ref{tab:rubber-results}, demonstrating the advantage of incorporating flexible representations of the strain energy function.
In contrast, models that use only {\emph{negative exponents}} $\alpha_-$, intended to reflect $I_2$-like behavior, perform poorly for rubber. These models fail to capture the dominant stretch-induced behavior and yield low accuracy, with average $R^2$ values as low as 0.195 for the FICNN and 0.264 for the GI-CANN. The corresponding visualizations show faint or absent blue regions, confirming the minimal contribution of negative $\alpha$ terms to the stress response.\\[4.pt]
\noindent For the brain data, a different mechanical behavior emerges. Human brain cortex, which deforms under much smaller strains and exhibits shear-dominant and pressure-sensitive characteristics, responds best to generalized invariants with large negative exponent values. These values correspond to an $I_2$-like behavior. Remarkably, even a single invariant $\mathcal{J}_{\alpha_-^*}$ with an optimally chosen negative exponent $\alpha_-^* = -18.016$ proves sufficient to accurately describe the tissue's response under all loading conditions, uniaxial tension, uniaxial compression, and previously unseen simple shear. The resulting one-term  strain energy function,
\begin{equation} \label{eq: best brain model}
    \psi = \mathrm{f_{id}}(\mathcal{J}_{-18.016}) = 0.009\,\mathrm{kPa} \cdot (\mathcal{J}_{-18.016} - 3),
\end{equation}
produces an average $R^2 = 0.966$, significantly surpassing traditional models built around $I_1$ or $I_2$. As shown in Figure~\ref{fig:brain-plots}, the third row visualizes the dominance of negative $\alpha_-$ contributions through strong blue regions in the stress maps. These patterns emphasize the model’s reliance on $I_2$-like deformation measures and the negligible role of any positiv $\alpha_+$ values, which are effectively pruned by the model’s sparsity-promoting regularization. Indeed, in the columns corresponding to $\alpha_+ > 0$, the red-shaded regions vanish, confirming the absence of positive $\alpha_+$ contributions. Among $2^3 -1$ possible isotropic models based on a single generalized invariant, our automatically discovered model in (\ref{eq: best brain model}) exactly aligns with findings in previous studies \cite{budday_mechanical_2017, mihai_comparison_2015,miller_mechanical_2000,mihai2017family} where the models have been manually selected. For example, among five traditional models, neo-Hookean, Mooney-Rivlin, Demiray, Gent, and Ogden, the one-term Ogden model with a high negative exponent was identified as the only model that best represents the response of human brain tissue under different loading conditions \cite{budday_mechanical_2017}. Moreover, the identified model parameters are close to the parameters recently identified via the  efficient unsupervised constitutive law identification EUCLID  \cite{flaschel_usupervised_2021,flaschel_automated_2023}.~Notably, the SI-CANN does not have the flexibility to discover the Ogden model for arbitrary exponent values, whereas the PS-CANN cannot identify models involving exponential or logarithmic functions. Our new GI-CANN inherently eliminates both these limitations.
\noindent The relative error plots in Figure~\ref{fig:error-plots-brain} show that the discovered model in (\ref{eq: best brain model}) again produces significantly smaller errors than the discovered \emph{standard-invariant-based} model $\psi(I_2)=\rm{f_{ln}}(I_2)$, which corresponds to the Gent model. Despite the small-strain regime and the complexity of soft tissue mechanics, these results slightly outperform, or at least match, those of the Anssari-Benam model  \cite{anssari-benam_large_2023}. The strength of the proposed method is again evident: It automatically discovers a well-suited model without \emph{a priori} assumptions about the strain energy function and naturally adapts to the tissue’s shear-dominant and pressure-sensitive response.

\noindent Due to the relatively small deformation during the brain experiments, the classical invariants $I_1$ and $I_2$ have nearly identical values. As a result, standard models based on either of these invariants produce similar fitting results. This behavior is reflected in the first and second rows of Figure~\ref{fig:brain-plots}, where the visualizations across the $I_1$- and $I_2$-like inputs appear nearly indistinguishable. Table~\ref{tab:brain-results} corroborates this finding, with models using $\alpha = 2$ or $\alpha = -2$ both yielding moderate $R^2$ values, but falling short of the high fidelity achieved with the optimized negative-exponent model. This observation aligns well with previous findings \cite{dammas_when_2025} and underscores the power of generalized invariants, significantly improving the goodness of fit compared to \emph{standard-invariant-based} models.
Although the FNN model achieves the highest average $R^2 = 0.982$ when combining both positive and negative invariants with fixed exponents, this result does not translate into a structurally interpretable formulation. Instead, the GI-CANN model demonstrates nearly the same accuracy with $R^2 = 0.966$, while maintaining interpretability and sparsity. The network naturally suppresses less relevant inputs, a fact confirmed by the absence of the red-shaded stress contributions in the third row of Figure~\ref{fig:brain-plots}. Moreover, the PS-CANN model with a dense range of discrete exponents from $-30$ to $30$ also performs competitively, but lacks the efficiency and interpretability of GI-CANN's learned invariant representations.\\[4.pt]
\noindent Taken together, our findings corroborate prior insights reported in the literature. For example, previous work suggested that the optimal exponents in \emph{generalized-invariant-based} models differ noticeably across different materials~\citep{anssari-benam_generalised_2024}, but those studies required a {\emph{two-step fitting}} process and full stress--stretch datasets. Our GI-CANN achieves comparable and often superior results with {\emph{single-step training}} and reduced data requirements. The model's predictive power, evaluated by the coefficient of determination \( R^2 \), consistently outperforms standard invariant neural models, constitutive neural networks \cite{linka_new_2023}, unconstrained feedforward neural networks, and fully input convex neural network \cite{dammas_when_2025} across all test configurations.\\


\begin{table}[ht]
\caption{Comparison of neural networks performance for rubber \cite{treloar_stress-strain_1944} with different neural network approaches and generalized invariants. Training is performed on uniaxial extensions (ut) and equibiaxial extensions (bt), while testing is conducted on pure shear (ps) data.} 
\label{tab:rubber-results}
\begin{flushleft}
\begin{tabularx}{\textwidth}{XX>{\raggedright\arraybackslash}m{3cm}ccc>{\centering\arraybackslash}X}
\toprule
\textbf{neural network type} & \textbf{invariants} & \textbf{exponent} & $R^2_{\mathrm{ut}}$ & $R^2_{\mathrm{bt}}$ & $R^2_{\mathrm{ps}}$ & \textbf{average $R^2$} \\
\midrule

\rowcolor{lightred!50}
SI-CANN \cite{linka_new_2023} & $\mathcal{J}_{\alpha_+}$ & $\alpha_+ = 2$ & 0.992 & 0.960 & 0.996 & 0.982 \\
\rowcolor{lightred!50}
FICNN \cite{dammas_when_2025} & $\mathcal{J}_{\alpha_+}$ & $\alpha_+ = 2$ & 0.987 & 0.966 & 0.954 & 0.969 \\
\rowcolor{lightred!50}
FNN \cite{dammas_when_2025} & $\mathcal{J}_{\alpha_+}$ & $\alpha_+ = 2$ & 0.983 & 0.984 & 0.968 & 0.978 \\
\rowcolor{lightred!50}
\textbf{GI-CANN} & $\mathcal{J}_{\alpha_+}$ & $\alpha^*_+ = 1.675$ & \textbf{0.999} & \textbf{0.997} & \textbf{0.998} & \textbf{0.998} \\

\rowcolor{lightblue!50}
SI-CANN \cite{linka_new_2023} & $\mathcal{J}_{\alpha_-}$ & $\alpha_- = -2$ & 0 & 0.954 & 0 & 0.318 \\
\rowcolor{lightblue!50}
FICNN \cite{dammas_when_2025} & $\mathcal{J}_{\alpha_-}$ & $\alpha_- = -2$ & 0 & 0.585 & 0 & 0.195 \\
\rowcolor{lightblue!50}
\textbf{FNN} \cite{dammas_when_2025} & $\mathcal{J}_{\alpha_-}$ & $\alpha_- = -2$ & \textbf{0} & \textbf{0.790} & \textbf{0.589} & \textbf{0.460} \\
\rowcolor{lightblue!50}
GI-CANN & $\mathcal{J}_{\alpha_-}$ & $\alpha^*_- = 0.155$ & 0.166 & 0 & 0.627 & 0.264 \\

\rowcolor{lightgray}
SI-CANN \cite{linka_new_2023} & $\mathcal{J}_{\alpha_+}, \mathcal{J}_{\alpha_-}$ & 
$\begin{aligned}
\alpha_+ & = 2 \\
\alpha_- & = -2
\end{aligned}$ 
& 0.994 & 0.984 & 0.952 & 0.977 \\

\rowcolor{lightgray}
FICNN \cite{dammas_when_2025} & $\mathcal{J}_{\alpha_+}, \mathcal{J}_{\alpha_-}$ & 
$\begin{aligned}
\alpha_+ & = 2 \\
\alpha_- & = -2
\end{aligned}$ 
& 0.996 & 0.991 & 0.993 & 0.993 \\

\rowcolor{lightgray}
FNN \cite{dammas_when_2025} & $\mathcal{J}_{\alpha_+}, \mathcal{J}_{\alpha_-}$ & 
$\begin{aligned}
\alpha_+ & = 2 \\
\alpha_- & = -2
\end{aligned}$ 
& 0.997 & 0.998 & 0.958 & 0.984 \\

\rowcolor{lightgray}
\textbf{GI-CANN} & $\mathcal{J}_{\alpha_+}, \mathcal{J}_{\alpha_-}$ & 
$\begin{aligned}
\alpha^*_+ & = 1.797 \\
\alpha^*_- & = -0.879
\end{aligned}$ 
& \textbf{0.998} & \textbf{0.999} & \textbf{1.000} & \textbf{0.999} \\
\bottomrule
\end{tabularx}
\end{flushleft}
\end{table}

\begin{table}[ht]
\caption{Comparison of neural network performance for brain tissue with different neural network approaches and generalized invariants. Training is performed on uniaxial extension (ut) and uniaxial compression (uc) data, whereas testing is performed on simple shear (ss) data.}
\label{tab:brain-results}
\begin{flushleft}
\begin{tabularx}{\textwidth}{XX>{\raggedright\arraybackslash}m{3.5cm}cccc}
\toprule
\textbf{neural network type} & \textbf{invariants} & \textbf{exponent} & $R^2_{\mathrm{ut}}$ & $R^2_{\mathrm{uc}}$ & $R^2_{\mathrm{ss}}$ & \textbf{average $R^2$} \\
\midrule

\rowcolor{lightred!50}
\textbf{SI-CANN} \cite{linka_new_2023} & $\mathcal{J}_{\alpha_+}$ & $\alpha_+ = 2$ & \textbf{0} & \textbf{0.706} & \textbf{0.875} & \textbf{0.527} \\
\rowcolor{lightred!50}
FICNN \cite{dammas_when_2025} & $\mathcal{J}_{\alpha_+}$ & $\alpha_+ = 2$ & 0 & 0.884 & 0 & 0.294 \\
\rowcolor{lightred!50}
FNN \cite{dammas_when_2025} & $\mathcal{J}_{\alpha_+}$ & $\alpha_+ = 2$ & 0 & 0.889 & 0 & 0.296 \\
\rowcolor{lightred!50}
GI-CANN & $\mathcal{J}_{\alpha_+}$ & $\alpha_+^* = 2.971$ & 0 & 0.700 & 0.879 & 0.526 \\

\rowcolor{lightblue!50}
SI-CANN \cite{linka_new_2023} & $\mathcal{J}_{\alpha_-}$ & $\alpha_- = -2$ & 0.120 & 0.836 & 0.879 & 0.610 \\
\rowcolor{lightblue!50}
FICNN \cite{dammas_when_2025} & $\mathcal{J}_{\alpha_-}$ & $\alpha_- = -2$ & 0.021 & 0.927 & 0.929 & 0.625 \\
\rowcolor{lightblue!50}
FNN \cite{dammas_when_2025} & $\mathcal{J}_{\alpha_-}$ & $\alpha_- = -2$ & 0.213 & 0.926 & 0.943 & 0.694 \\
\rowcolor{lightblue!50}
\textbf{GI-CANN} & $\mathcal{J}_{\alpha_-}$ & $\alpha^*_- = -18.016$ & \textbf{0.930} & \textbf{0.998} & \textbf{0.970} & \textbf{0.966} \\

\rowcolor{lightgray}
SI-CANN \cite{linka_new_2023} & $\mathcal{J}_{\alpha_+}, \mathcal{J}_{\alpha_-}$ & 
$\begin{aligned}
\alpha_+ &= 2 \\
\alpha_- &= -2
\end{aligned}$ & 0.140 & 0.832 & 0.878 & 0.617 \\

\rowcolor{lightgray}
FICNN \cite{dammas_when_2025} & $\mathcal{J}_{\alpha_+}, \mathcal{J}_{\alpha_-}$ & 
$\begin{aligned}
\alpha_+ &= 2 \\
\alpha_- &= -2
\end{aligned}$ & 0.368 & 0.948 & 0 & 0.439 \\

\rowcolor{lightgray}
\textbf{FNN} \cite{dammas_when_2025} & $\mathcal{J}_{\alpha_+}, \mathcal{J}_{\alpha_-}$ & 
$\begin{aligned}
\alpha_+ &= 2 \\
\alpha_- &= -2
\end{aligned}$ & \textbf{0.999} & \textbf{1.000} & \textbf{0.948} & \textbf{0.982}\\

\rowcolor{lightgray}
PS-CANN \cite{pierre_principal-stretch-based_2023} & $\mathcal{J}_{-30}, \ldots, \mathcal{J}_{30}$ & 
$\alpha \in \{-30,\ldots,30\}$ & 0.926 & 0.998 & 0.988 & 0.974 \\

\rowcolor{lightgray}
GI-CANN & $\mathcal{J}_{\alpha_+}, \mathcal{J}_{\alpha_-}$ & 
$\begin{aligned}
\alpha_+^* &= 0.741 \\
\alpha_-^* &= -18.223
\end{aligned}$ & 0.930 & 0.997 & 0.972 & 0.966 \\

\bottomrule
\end{tabularx}
\end{flushleft}
\end{table}

\section{Conclusion and Outlook}
 
\noindent 
In this study, we propose a new class of constitutive neural networks that integrates \emph{standard-invariant-based} and \emph{principal-stretch-based} constitutive artificial neural networks as special cases:~\emph{generalized-invariant-based} constitutive artificial neural networks or GI-CANNs.
We demonstrate the performance of our new GI-CANNs using two complementary datasets for rubber and for human brain tissue. 
Our results highlight the versatility and robustness of the proposed GI-CANN framework in discovering interpretable and accurate constitutive models across different material classes. 
By relying on generalized invariants \( \mathcal{J}_\alpha \) \cite{anssari-benam_generalised_2024}, 
GI-CANN circumvents the limitations of previous approaches that require an \emph{a priori} specification of the  strain energy function  and manual selection of relevant invariants. Instead, both the structure of the invariants and the form of the strain energy function are discovered directly from data. This approach proves particularly effective for soft matter, such as rubber and brain tissue.

\noindent Our findings align well with the trends of recent studies that also reported the advantage of a flexible invariant design in capturing the complex behavior of soft materials \cite{kumar_tube_2023,dammas_when_2025}. Our results extend this observation by directly comparing different classes of neural network formulations and by optimizing the invariant exponents to adapt to the material-specific responses.

\noindent For rubber, our findings confirm that the optimal invariant structure is closely aligned with the classical first invariant \( I_1 \), consistent with well-established models including the neo-Hookean or Mooney–Rivlin formulations, that reflect the general notion that rubber exhibits an entropic elasticity, meaning that the majority of its elastic response arises from changes in chain conformational entropy, rather than from changes in internal energy \cite{treloar1975physics}. At the macrosocopic level, the first invariant captures the average chain extension \cite{kuhl_i_2024} and the elastic energy in rubber is mostly stored through entropic effects related to molecular chain extension \cite{treloar1975physics}. The best-performing GI-CANN models feature positive exponents \( \alpha_+ \), suggesting that stretch-dominant behaviors are well captured by \( \mathcal{J}_{\alpha_+} \)-type invariants. For brain tissue, in contrast, the data are better represented using \( \mathcal{J}_{\alpha_-} \) with large negative exponents. This suggests that the molecular origin of elasticity in brain tissue is fundamentally different than in rubber. 

\noindent Despite these promising results, several limitations remain, opening the door for further investigations. First, our current implementation is restricted to isotropic materials and assumes incompressibility. Extending the GI-CANN to anisotropic or compressible formulations would require careful adaptation of the invariant structures and associated input features. Second, although the GI-CANN discovers optimal exponents \( \alpha^* \) during training, the physical interpretation of highly negative values can be challenging \cite{anssari-benam_modelling_2022}. However, similar exponents have been reported in previous studies \cite{budday_mechanical_2017, mihai_comparison_2015,miller_mechanical_2000,mihai2017family,flaschel_automated_2023}, indicating that such values are commonly used to capture nonlinear material behavior. While these exponents improve predictive accuracy, their mechanistic meaning is still partially open, and future work could incorporate physics-informed constraints \cite{Raissi2019} or Bayesian priors \cite{linka_2025} to guide \(\alpha\) toward more interpretable ranges. Third, the training process, while automated, still requires hyperparameter tuning of both the network architecture and the regularization process. The current choices were selected empirically based on validation performance and motivated by previous works to ensure comparability. Finally, while GI-CANN performs well even with limited experimental data, the accuracy of discovered models depends on the quality and diversity of the training data.  In particular, if the loading conditions fail to sufficiently excite the relevant deformation modes, the learned invariants may not generalize. This issue is especially pertinent for biological tissues, where experimental datasets are often sparse or noisy. However, at present, no universal protocol exists, but we chose the deformation modes reported here to ensure comparability with prior studies \cite{linka_new_2023,linka_automated_2023-1,dammas_when_2025}.\\[4.pt]
\noindent Overall, \emph{generalized-invariant-based} constitutive artificial neural networks provide a compelling tool to discover interpretable models for soft matter systems by combining physical insight with data-driven flexibility. Future work will explore extensions to time-dependent behavior, anisotropy, and uncertainty quantification to broaden the applicability, robustness, and adaptation of the approach.

\bibliographystyle{elsarticle-harv}

\section*{Data availability}
\noindent
Our source code, data, and examples are available at https://github.com/LivingMatterLab/CANN.

\section*{Acknowledgments}
\noindent
The authors acknowledge support from the 
National Science Foundation (NSF) Grant CMMI 2320933 and the
European Research Council (ERC) Grant 101141626 DISCOVER. Funded by the European Union. Views and opinions expressed are however those of the authors only and do not necessarily reflect those of the European Union or the European Research Council
Executive Agency. Neither the European Union nor the granting authority can be held responsible for them. The authors utilized ChatGPT and HAWKI, a large language model interface provided by Friedrich-Alexander-Universität Erlangen-Nürnberg, to enhance the writing style in certain sections of the manuscript. After using these tools, the authors reviewed and edited the content as needed and take full responsibility for the content of the publication. 
\newpage
\appendix
\section{Experimental stress-strain and stress-stretch data}
\begin{table}[!ht]
\centering
\caption{{\bf{\sffamily{Cauchy stress data for rubber and brain cortex under various deformation modes.}}} \label{tab:rubber_brain_data}
Rubber is tested in uniaxial tension (ut), equibiaxial tension (bt), and pure shear (ps), with stress in MPa \cite{treloar_stress-strain_1944} Brain cortex is tested in uniaxial tension (ut), uniaxial compression (uc), and simple shear (ss), with stress in kPa \cite{budday_mechanical_2017}.
}
\vspace{0.2cm}
\small
\renewcommand{\arraystretch}{0.9}
\begin{tabular}{|c|c||c|c||c|c||c|c||c|c||c|c|}
\hline
\multicolumn{6}{|c||}{\bf{\sffamily{rubber (MPa)}}} & \multicolumn{6}{c|}{\bf{\sffamily{brain cortex (kPa)}}} \\ \hline \hline
\multicolumn{2}{|c||}{\textbf{ut}} & 
\multicolumn{2}{c||}{\textbf{bt}} & 
\multicolumn{2}{c||}{\textbf{ps}} & 
\multicolumn{2}{c||}{\textbf{ut}} & 
\multicolumn{2}{c||}{\textbf{uc}} & 
\multicolumn{2}{c|}{\textbf{ss}} \\ \hline
\textbf{$\lambda_1$} & \textbf{$\sigma_{11}$} & 
\textbf{$\lambda_1$} & \textbf{$\sigma_{11}$} & 
\textbf{$\lambda_1$} & \textbf{$\sigma_{11}$} & 
\textbf{$\lambda_1$} & \textbf{$\sigma_{11}$} & 
\textbf{$\lambda_1$} & \textbf{$\sigma_{11}$} & 
\textbf{$\gamma$} & \textbf{$\sigma_{12}$} \\ \hline
1.00 & 0.00  & 1.00 & 0.00   & 1.00 & 0.00   & 1.00 & 0.00  & 1.00 & 0.00   & 0.00 & 0.00 \\
1.13 & 0.15  & 1.08 & 0.17   & 1.05 & 0.10   & 1.01 & 0.03  & 0.99 & -0.03  & 0.01 & 0.01 \\
1.41 & 0.47  & 1.15 & 0.26   & 1.13 & 0.20   & 1.01 & 0.05  & 0.98 & -0.08  & 0.03 & 0.03 \\
1.89 & 0.98  & 1.21 & 0.40   & 1.20 & 0.30   & 1.02 & 0.07  & 0.97 & -0.13  & 0.04 & 0.05 \\
2.45 & 1.67  & 1.32 & 0.58   & 1.33 & 0.40   & 1.03 & 0.08  & 0.97 & -0.18  & 0.05 & 0.06 \\
3.06 & 2.65  & 1.43 & 0.74   & 1.45 & 0.60   & 1.03 & 0.10  & 0.96 & -0.23  & 0.06 & 0.08 \\
3.62 & 3.83  & 1.70 & 1.12   & 1.86 & 1.10   & 1.04 & 0.12  & 0.96 & -0.28  & 0.08 & 0.13 \\
4.06 & 5.03  & 1.95 & 1.51   & 2.40 & 1.80   & 1.05 & 0.16  & 0.95 & -0.33  & 0.10 & 0.15 \\
4.82 & 7.71  & 2.50 & 2.42   & 2.99 & 2.80   & 1.06 & 0.19  & 0.94 & -0.40  & 0.11 & 0.18 \\
5.41 & 10.50 & 3.04 & 3.82   & 3.50 & 4.00   & 1.06 & 0.21  & 0.94 & -0.47  & 0.13 & 0.22 \\
5.79 & 13.30 & 3.44 & 5.07   & 3.98 & 5.20   & 1.07 & 0.24  & 0.93 & -0.53  & 0.14 & 0.26 \\
6.23 & 16.70 & 4.03 & 7.95   & 4.39 & 6.50   & 1.08 & 0.27  & 0.92 & -0.61  & 0.15 & 0.31 \\
6.96 & 26.40 & 4.26 & 9.50   & 4.72 & 7.80   & 1.09 & 0.33  & 0.91 & -0.70  & 0.16 & 0.37 \\
7.25 & 32.60 & 4.45 & 10.90  & 4.99 & 9.10   & 1.09 & 0.37  & 0.91 & -0.80  & 0.18 & 0.45 \\
--   & --    & --   & --     & --   & --     & 1.10 & 0.45  & 0.90 & -1.03  & 0.20 & 0.54 \\
\hline
\end{tabular}
\end{table}

\section{Discovered model parameters}

\begin{table}[!ht]
\centering
\caption{{\bf{\sffamily{Model specification for rubber.}}} Discovered material parameters for SI-CANN and GI-CANN in the general strain energy function in (\ref{eq:general free energy}). The corresponding values of the exponents $\alpha$ and goodness of fit are given in Table \ref{tab:rubber-results}.} \label{tab:rubber_parameters}
\vspace{0.2cm}
\small
\renewcommand{\arraystretch}{1.2}
\resizebox{\textwidth}{!}{%
\begin{tabular}{|l|l||c|c||c|c||c|c|}
\hline
{\vtop{\hbox{\strut\bf{\sffamily{network weights}}}}} &
\multicolumn{1}{l||}{\vtop{\hbox{\strut\bf{\sffamily{model term}}}}} &
\multicolumn{1}{c|}{\vtop{\hbox{\strut\bf{\sffamily{SI-CANN}}}\hbox{\strut$\mathcal{S} = \{2\}$}}} &
\multicolumn{1}{c||}{\vtop{\hbox{\strut\bf{\sffamily{GI-CANN}}}\hbox{\strut$\mathcal{S} = \{\mathbb R ^+\}$}}} &
\multicolumn{1}{c|}{\vtop{\hbox{\strut\bf{\sffamily{SI-CANN}}}\hbox{\strut$\mathcal{S} = \{-2\}$}}} &
\multicolumn{1}{c||}{\vtop{\hbox{\strut\bf{\sffamily{GI-CANN}}}\hbox{\strut$\mathcal{S} = \{\mathbb R ^-\}$}}} &
\multicolumn{1}{c|}{\vtop{\hbox{\strut\bf{\sffamily{SI-CANN}}}\hbox{\strut$\mathcal{S} = \{-2,2\}$}}} &
\multicolumn{1}{c|}{\vtop{\hbox{\strut\bf{\sffamily{GI-CANN}}}\hbox{\strut$\mathcal{S} = \{\mathbb R\}$}}} \\ \hline
$w^{\mathrm{id}}_{1}$[MPa] & $[\mathcal{J}_{\alpha_+} - 3]$ 
& 0.137 & 0.227 & - & - & - & 0.175 \\ \hline
$w^{\mathrm{id}}_{2}$ [MPa]& $[\mathcal{J}_{\alpha_-} - 3]$ 
& - & - & 0.015 & 108.744 & 0.015 & 0.025 \\ \hline
$w^{\mathrm{exp}}_{1,1} [-],\; w^{\mathrm{exp}}_{1,2}$ [MPa] & $\exp([\mathcal{J}_{\alpha_+} - 3]) - 1$ 
& 0.048, 0.355 & 0.107, 0.297 & - & - & 0.007, 0.132 & - \\ \hline
$w^{\mathrm{exp}}_{2,1} [-],\; w^{\mathrm{exp}}_{2,2} $ [MPa] & $\exp([\mathcal{J}_{\alpha_-} - 3]) - 1$  
& - & - & - & - & 0.050, 0.057 & 0.304, 0.303 \\ \hline
$w^{\mathrm{ln}}_{1,1} [-],\; w^{\mathrm{ln}}_{1,2}$ [MPa]& $\ln(1 - [\mathcal{J}_{\alpha_+} - 3])$ 
& - & - & - & - & - & - \\ \hline
$w^{\mathrm{ln}}_{2,1} [-],\; w^{\mathrm{ln}}_{2,2}$ [MPa]& $\ln(1 - [\mathcal{J}_{\alpha_-} - 3])$ 
& - & - & - & - & - & - \\ \hline
\end{tabular}
}
\end{table}

\begin{table}[!ht]
\centering
\caption{{\bf{\sffamily{Model specification for brain cortex.}}} Discovered material parameters for SI-CANN and GI-CANN in the general strain energy function (\ref{eq:general free energy}). The corresponding values of the exponents $\alpha$ and goodness of fit are given in Table \ref{tab:brain-results}.} \label{tab:brain_parameters}
\vspace{0.2cm}
\small
\renewcommand{\arraystretch}{1.2}
\resizebox{\textwidth}{!}{%
\begin{tabular}{|l|l||c|c||c|c||c|c|}
\hline
{\vtop{\hbox{\strut\bf{\sffamily{network weights}}}}} &
\multicolumn{1}{l||}{\vtop{\hbox{\strut\bf{\sffamily{model term}}}}} &
\multicolumn{1}{c|}{\vtop{\hbox{\strut\bf{\sffamily{SI-CANN}}}\hbox{\strut$\mathcal{S} = \{2\}$}}} &
\multicolumn{1}{c||}{\vtop{\hbox{\strut\bf{\sffamily{GI-CANN}}}\hbox{\strut$\mathcal{S} = \{\mathbb R ^+\}$}}} &
\multicolumn{1}{c|}{\vtop{\hbox{\strut\bf{\sffamily{SI-CANN}}}\hbox{\strut$\mathcal{S} = \{-2\}$}}} &
\multicolumn{1}{c||}{\vtop{\hbox{\strut\bf{\sffamily{GI-CANN}}}\hbox{\strut$\mathcal{S} = \{\mathbb R ^-\}$}}} &
\multicolumn{1}{c|}{\vtop{\hbox{\strut\bf{\sffamily{SI-CANN}}}\hbox{\strut$\mathcal{S} = \{-2,2\}$}}} &
\multicolumn{1}{c|}{\vtop{\hbox{\strut\bf{\sffamily{GI-CANN}}}\hbox{\strut$\mathcal{S} = \{\mathbb R\}$}}} \\ \hline
$w^{\mathrm{id}}_{1}$[kPa] & $[\mathcal{J}_{\alpha_+} - 3]$ 
& 0.997 & - & - & - & - & - \\ \hline
$w^{\mathrm{id}}_{2}$ [kPa]& $[\mathcal{J}_{\alpha_-} - 3]$ 
& - & - & - & 0.009 & - & - \\ \hline
$w^{\mathrm{exp}}_{1,1} [-],\; w^{\mathrm{exp}}_{1,2}$ [kPa] & $\exp([\mathcal{J}_{\alpha_+} - 3]) - 1$ 
& - & 0.703, 0.626 & - & - & - & - \\ \hline
$w^{\mathrm{exp}}_{2,1} [-],\; w^{\mathrm{exp}}_{2,2} $ [kPa] & $\exp([\mathcal{J}_{\alpha_-} - 3]) - 1$  
& - & - & - & - & - & 0.096, 0.096 \\ \hline
$w^{\mathrm{ln}}_{1,1} [-],\; w^{\mathrm{ln}}_{1,2}$ [kPa]& $\ln(1 - [\mathcal{J}_{\alpha_+} - 3])$ 
& - & - & - & - & - & - \\ \hline
$w^{\mathrm{ln}}_{2,1} [-],\; w^{\mathrm{ln}}_{2,2}$ [kPa]& $\ln(1 - [\mathcal{J}_{\alpha_-} - 3])$ 
& - & - & 1.090, 0.921 & - & 1.064, 0.941 & - \\ \hline
\end{tabular}
}
\end{table}
\end{document}